\documentclass[twocolumn, aps, showpacs, epsf]{revtex4}
\usepackage{graphicx}
\usepackage{dcolumn}
\usepackage{amsmath}
\usepackage{amssymb}

\usepackage[colorlinks]{hyperref}
\hypersetup{citecolor=blue}

\usepackage{bm}
\usepackage{epstopdf}
\usepackage{amsthm}
\usepackage{float}
\graphicspath{{./figures/}}

\def\ie{i.e., }

\def\ep{\varepsilon}

\def\ie{i.e. }

\def\beqr{\begin{eqnarray}}
\def\eqnr{\end{eqnarray}}
\def\beq{\begin{equation}}
\def\bc{\begin{center}}
\def\ec{\end{center}}
\def\eqn{\end{equation}\noindent}
\begin{document}

\title{Explosive transitions in interacting star networks}

\author{Anjuman Ara Khatun$^1$, Ruby Varshney$^1$ and, Haider Hasan Jafri$^1$ }

\affiliation{$^1$Department of Physics, Aligarh Muslim University, Aligarh 202 002, India}

\begin{abstract}

We study transition to phase synchronization in an ensemble of Stuart-Landau oscillators interacting on a star network. We observe that by introducing frequency weighted coupling and time scale variations in the dynamics of nodes, system exhibits a first order explosive transition to phase synchrony. Further, we extend this study to understand the nature of synchronization in case of two coupled star networks. In presence of symmetry preserving (direct) coupling between the hubs of the two stars, we observe that hysteresis width first increases and then saturates for increasing inter-star coupling strength. For symmetry breaking (conjugate) coupling, the hysteresis width
first increases and then decreases with increasing inter-star coupling. As we increase the inter-star
coupling further, the transition gradually becomes a second order.
\end{abstract}
\maketitle

\section{Introduction} \label{sec:intro}

Synchronization is an important phenomenon in the studies devoted to emergence of collective behaviour in ensembles of networked chaotic systems. In such a scenario, collective dynamics may be synchronized, desynchronized, clusters or a chimera state \cite{pikovsky-cup-2001}. Numerous studies have been devoted to the study of these phenomenon in an ensemble of chaotic oscillators. Synchronization being ubiquitous in the real world, has been studied extensively in complex networks \cite{boccaletti-pr-2006, arenas-pr-2008, dorogovtsev-rmp-2008}. Such an onset of coherence in real systems have been reported in power grids, neuronal networks, communication networks, and circadian rhythm. Most studies suggest that transitions to synchrony are second order where the order parameter that distinguishes synchrony from desynchrony varies smoothly with the coupling strength. Recently, it has been reported that a discontinous transition namely explosive synchronization is possible in a network of coupled periodic oscillators. This transition was shown to occur on a scale free network as a result of positive correlations between the node degree and the natural frequency of the oscillator \cite{gardenes-prl-2011}.

Since the discovery of explosive synchronization (ES) transition in an ensemble of coupled oscillators, phenomenon of phase transition has attracted lot of attention \cite{gardenes-prl-2011}. Explosive synchonization is characterized by an abrupt jump in the global order parameter along with a hysteresis loop as the coupling is increased. It was shown that by incorporating time delay in a network of oscillators having frequency-degree correlation, it was possible to enhance ES to reach a synchronous state \cite{peron-pre-2012}. Certain studies have shown that it is also possible to obtain ES by considering a second order Kuramoto oscillator where inertia plays an important role \cite{tanaka-prl-1997}. Cluster explosive synchronization characterized by a cascade of transition towards a synchronous state has been reported in cases where node dynamics is on a second order Kuramoto oscillator \cite{ji-prl-2013}. Recent studies have shown that ES can also be observed in multilayer networks under various conditions namely inter-layer coupling \cite{zhang-prl-2015, danziger-np-2019}, inertia \cite{kachhvah-epl-2017}, time delay and inhibitory layers \cite{jalan-pre-2019},  etc.

However, the studies are mainly restricted to phase oscillators \cite{gardenes-prl-2011, zhang-pre-2013, coutinho-pre-2013} except a few cases where this transition has been observed in coupled oscillators having amplitudes \cite{leyva-prl-2012, chen-chaos-2013, boaretto-pre-2019, sharma-pl-2019}. In real world systems, the nodes can have dynamics that may be complicated and involve amplitudes along with the phase. Therefore, a number of studies have considered coupled Stuart-Landau (SL) oscillators since it is a good model for the dynamics in the vicinity of supercritical Hopf bifurcation. This model also provides a rotational invariance that may be preserved or destroyed in presence of interactions \cite{punetha-pre-2018}. Thus, SL oscillator is a useful model to study the collective dynamics of a network of oscillators coupled in such a way that the symmetry is either preserved or destroyed \cite{premalatha-pre-2015, schneider-pre-2015, zakharova-jp-2016}.

In heterogenous networks, hubs play a central role in deciding the emergent dynamics. Numerous studies have argued that star network is a useful model to understand important aspect of a heterogenous network. However, there are situations where a heterogenous network may have multiple hubs. Such a situation may be modelled by considering a configuration made up of two (or more) coupled star units. It has been shown that the most effective technique to couple two stars is through their hubs \cite{aguirre-np-2013, aguirre-prl-2014}.

In the present work, our interest is to explore the transition to synchrony in networks of Stuart-Landau  oscillators on a star graph. We consider a very general coupling scheme where the oscillators are coupled under the influence of frequency weighted coupling \cite{zhang-pre-2013}. Further, to ensure positive correlation between the natural frequency and the oscillator degree we introduce the time scale parameters so that natural frequencies of the oscillators can be adjusted \cite{leyva-prl-2012, pyragas-pre-1996}. Under these settings, we show that a heterogenous network of SL oscillators exhibit first order ES. Real networks often interact with other networks and it was shown that connecting highest degree nodes is the most effective strategy to influence the two units \cite{aguirre-np-2013, aguirre-prl-2014}. Therefore, we extend the analysis done on a single star to a case where the two identical star networks are coupled through their hubs. Through extensive numerical simulations, we outline techniques to avoid or enhance various explosive transitions. We observe that if the coupling is done in such a way that it preserves symmetry, hysteresis width increases . However, for the symmetry breaking interactions, we observe that for small values of inter-star couplings, the transition is a first order in nature with decreasing hysteresis width. As the coupling is increased further, we observe that the transition eventually becomes a second order transition.

The paper is organized as follows. In Sec.~ \ref{sec-II}, we discuss phase transitions in star network in presence of frequency weighted coupling and time scale parameter. In Sec. \ref{coupled}, we study the dynamics of coupled star networks interacting through direct diffusive coupling and conjugate coupling. Summary of the important results is presented in Sec.~\ref{summary}.

\section{STUART-LANDAU OSCILLATORS ON A STAR NETWORK} \label{sec-II}
\begin{figure} 
\centering
\includegraphics [scale=0.45,angle=0]{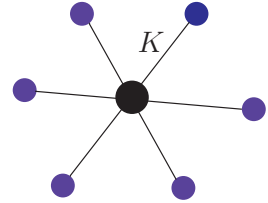}
\caption{Schematic view of a star network where six leaf nodes are connected to a central hub. In our model, the natural frequencies of the leaf nodes are all equal to $\omega_j$ (drawn from a Gaussian distribution $[0.5 : 1.5]$,) and the natural frequency of the hub is $\alpha \omega_h$, where $\omega_h=2$ is fixed and $\alpha$ is varied. Here, each leaf interacts with the hub through coupling strength $K$ and the hub experiences with the averaged coupling with intensity $\alpha K$.}
\label{fig1}
\end{figure}

Here we investigate the behaviour of phase transition in an ensemble of Stuart-Landau oscillator units on a star network via a bidirectional diffusive coupling. We consider a general framework by introducing a positive correlation between the coupling strength and the absolute of their natural frequency as suggested in Refs. \cite{zhang-pre-2013, bi-epl-2014}. Consider an ensemble of $N$ coupled SL oscillators with linear diffusive coupling (See Fig.~\ref{fig1} for illustration). The dynamical equations for the leaves are given by

\beq
\label{eq:system-eq1}
\dot{z}_{j} = \left[ (q+i\omega_j-|z_j|^2)z_j(t) + K |\omega_j|  \left(z_h(t)-z_j(t)\right )\right] 
\eqn

and the dynamics of the hub is given by

\begin{equation}
 \label{eq:system-eq1-hub}
\dot{z}_{h} = \alpha\left[ (q+i\omega_h-|z_h|^2)z_h(t) + \frac{K |\omega_h|}{N} \sum_{j=1}^{N}  \left(z_j(t)-z_h(t)\right )\right], 
\end{equation}
\noindent
where, $j = 1, \dots, N$ label the oscillators on the leaves and
subscript $h$ describes the variables of the hub. $z_j(t) = x_j(t) + iy_j(t)$ is the complex amplitude of the $j-$th oscillator at time $t$. For $q > 0$, the dynamics settles on a limit cycle while it is a fixed point for $q < 0$. Natural frequency of each oscillator is given by $\omega_j$ and $K$ is the coupling strength. Note that the coupling strength $K$ is multiplied by $|\omega_j|$ to incorporate frequency weighted coupling. Throughout the study, $\omega_j$ is drawn from a
Gaussian distribution $[0.5 : 1.5]$, $\omega_h = 2$ and $q = 1$. Time scale parameter $\alpha$ has been introduced to adjust the natural frequency of the oscillator on the hub as described in \cite{leyva-prl-2012, pyragas-pre-1996}. In this way it is possible to introduce degree-frequency correlation by setting $\alpha > 1$.

For SL oscillators, phase can be defined in $x-y$ plane and its instantaneous value for the $i-$th oscillator is given by $\phi_i (t) =$ arctan$[y_i(t)/x_i(t)]$. Mean synchronization degree is measured by defining a global order parameter
\beq
R=\left \langle \left |\frac{1}{N}\sum_{j=1}^N e^{i \phi_j(t)} \right|\right\rangle  _t
\label{eq-r}
\eqn 
\noindent
where $|.|$ denotes the modulus and $\langle . \rangle_t$ is used to describe averaging over time. For $R \sim 0$, phase of the oscillators are distributed uniformly over a unit circle and the state is desynchronized whereas for large values of $R~(R \sim 1)$ there exists exact phase synchronization. Throughout the study, the equations of the coupled oscillators are solved using the fourth–order Runge–Kutta routine with step size $dt = 0.01$. In both forward and backward calculations, the order parameter $R$ is evaluated by changing the coupling strength adiabatically. For forward transitions, we start with random initial conditions, and then the value of the coupling parameter is gradually increased such that the final state becomes the initial condition for the next value of coupling parameter. For the backward transition, we start from a coherent state and decrease the coupling strength adiabatically. For each coupling value, we start from initial conditions that are very close to the fully coherent state. 

\begin{figure} 
\includegraphics [scale=0.30,angle=270]{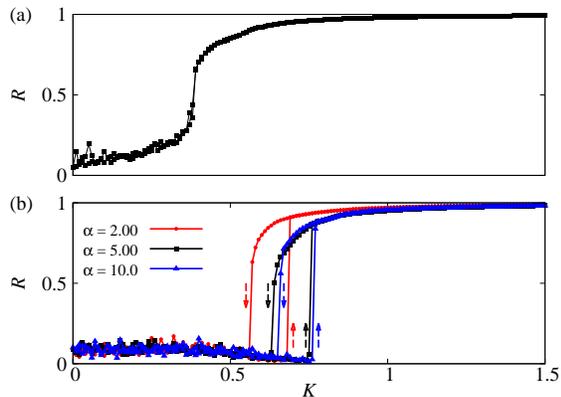}
\caption{Variation of the order parameter $R$ for Eqs.~\ref{eq:system-eq1} and \ref{eq:system-eq1-hub} for $N=500$ oscillators (a) without introducing time scale variations $(\alpha = 1)$ and (b) with time scale variations in the hub for $\alpha = 2, 5, 10$ as shown.}
\label{fig2}
\end{figure}

\begin{figure} 
\includegraphics [scale=0.30,angle=270]{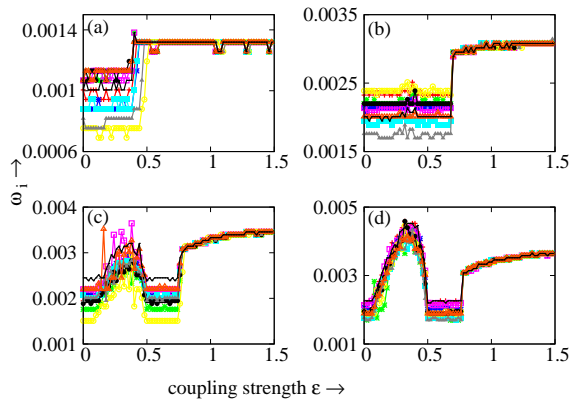}
\caption{Variation of frequency of $10$ different oscillators of the ensemble (Eqs.~\ref{eq:system-eq1} and \ref{eq:system-eq1-hub}) for (a) $\alpha = 1$, (b) $\alpha = 2$, (c) $\alpha = 5$ and (d) $\alpha = 10$.}
\label{fig3}
\end{figure}

In Fig.~\ref{fig2} order parameter $R$ for the ensemble (Eqs.~\ref{eq:system-eq1} and \ref{eq:system-eq1-hub}) of $N=500$ oscillators are represented. As shown in Fig.~\ref{fig2}(a), in the absence of time scale variations $(\alpha = 1)$, the order parameter $R$ shows an abrupt transition with a very small hysteresis. However, if we introduce time scale variations, we observe that the order parameter changes discontinuously  with a well defined hysteresis indicating a first order transition to synchrony. As shown in Fig.~\ref{fig2}(b), for different values of $\alpha$, we observe two sharp transitions namely  the forward and the backward continuation. This suggests that the system exhibits an explosive synchronization. Corresponding frequencies for each value of $\alpha$ have been plotted in Fig.~\ref{fig3} where we plot the variation of frequencies for $10$ oscillators.

\begin{figure}
\centering
\scalebox{0.35}{\includegraphics[angle=270]{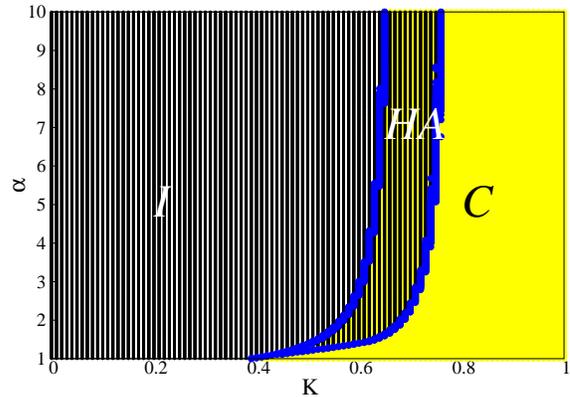}}
\caption{Dynamics in $\alpha- \ep$ plane showing region of incoherence $(I)$ in black colour, hysteresis area $(HA)$ in overlap region (black-yellow) and yellow region indicates coherent states $(C)$.}
\label{fig4}
\end{figure}

To explore the effect of time scale variations, we describe the transitions in $\alpha - K$ parameter space as shown in Fig.~\ref{fig4}. Region in black and yellow colour represents incoherent (I) and coherent (C) regimes respectively, whereas the overlap (black-yellow) region between two blue lines is the region of hysteresis area marked as HA. For $\alpha = 1$, the hysteresis region is almost negligible. This observation is consistent with the one observed in Fig.~\ref{fig2}(a). As the time scale variation $\alpha$ is increased, we observe that the first order transition with a well defined hysteresis region is obtained. As $\alpha$ is increased further, we observe that the hysteresis area (HA) first increases and then saturates as shown by the region between the two blue lines in Fig.~\ref{fig4}. This observation is consistent with the results as shown in Fig.~\ref{fig2}(b) which is plotted for $\alpha = 2, 5, 10$. \newline

\section{Coupled star networks}

\label{coupled}
\begin{figure} 
\centering
\includegraphics [scale=0.45,angle=0]{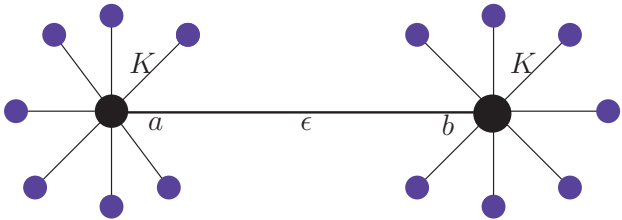}
\caption{Schematic view of the two coupled star network. Here, hubs (solid black circles) named as $a$ and $b$ are connected through the coupling strength $\varepsilon$. $K$ is the interaction strength between leaf nodes and the hub for each star network.}
\label{fig5}
\end{figure}

In heterogenous networks, hubs play an important role in defining the emergent properties. A single star network may be  helpful in understanding some of the important properties \cite{coutinho-pre-2013, peron-pre-2012, zou-prl-2014, perira-prl-2013, vlasov-pre-2015} but there are instances where it is possible to have  multiple hubs. Therefore, we design a heterogenous network where there may be two hubs that are interacting with each other. The emergent dynamics can be explained in terms of the collective dynamics of individual stars. Fig.~\ref{fig5} represents schematic sketch of two star network connected through their hubs. If two stars are denoted by $a$ and $b$, then, one can track the phase transition in each star by means of order parameters $R_a$ and $R_b$ defined as:
\beqr
R_a e^{i \psi_a}&=& \frac{1}{N} \sum_{j=1}^Ne^{i \theta_{j,a}}\\
R_b e^{i \psi_b}&=& \frac{1}{N} \sum_{j=1}^Ne^{i \theta_{j,b}}
\eqnr
where $\psi_a$ and $\psi_b$ are the average phases of the star networks $a$ and $b$, respectively. $\theta_{j,a}$ and $\theta_{j,b}$ are the phases of the SL oscillators on the nodes of individual star networks. Note that the technique to calculate the phase for SL has been outlined in Sec.~\ref{sec-II}. The global dynamics is described by a global order parameter defined as
\beq
R e^{i \psi}= \frac{1}{2N} \sum_{j=1}^{2N}e^{i \theta_{j}},
\eqn
where, the value of $\psi$ accounts for the average phase of the collective dynamics of the whole network and $\theta_j$ represents the phase with $j=1, \dots, 2N$.

\subsection{Direct diffusive coupling}

\begin{figure}
\centering
\scalebox{0.35}{\includegraphics[angle=270]{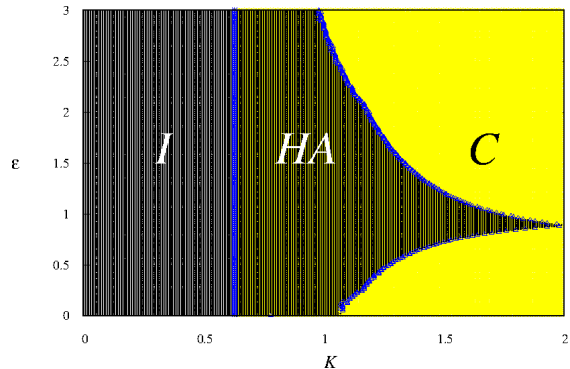}}
\caption{Global dynamics of the two coupled stars given by Eq.~\ref{eq-dif} in $K-\ep$ space. Black and yellow regions denoted by $I$ and $C$ indicate incoherent and coherent states. Region in between the blue lines represents the hysteresis area and is identified as HA.}
\label{fig6}
\end{figure}

We consider two star networks interacting with each other through their hubs via direct diffusive coupling. Dynamical equations for this case are given by 
%
\begin{widetext}
\[
\begin{aligned}
\label{eq-dif}
\dot{z}_{a,j} &= \left[ (q+i\omega_{a,j}-|z_{a,j}|^2)z_{a,j}(t)+ K |\omega_{a,j}|  \left(z_{a,h}(t)-z_{a,j}(t)\right )\right], \nonumber\\
\dot{z}_{a,h} &= \alpha \bigl\{ (q+i\omega_{a,h}-|z_{a,h}|^2)z_{a,h}(t) +\frac{K |\omega_{a,h}|}{N} \sum_{j=1}^{N}  \left(z_{a,j}(t)-z_{a,h}(t)\right ) + \ep \left({z}_{b,h}(t)-z_{a,h}(t)\right)\bigr\}, \nonumber   \\
\dot{z}_{b,j} &= \left[ (q+i\omega_{b,j}-|z_{b,j}|^2)z_{b,j}(t) + K |\omega_{b,j}|  \left(z_{b,h}(t)-z_{b,j}(t)\right )\right], \nonumber \\
\dot{z}_{b,h} &= \alpha \bigl\{ (q+i\omega_{b,h}-|z_{b,h}|^2)z_{b,h}(t)  +\frac{K |\omega_{b,h}|}{N} \sum_{j=1}^{N}  \left(z_{b,j}(t)-z_{b,h}(t)\right ) + \ep \left({z}_{a,h}(t)-z_{b,h}(t)\right )\bigr\}, 
\end{aligned}
\]    
\end{widetext}


where, subscripts $a$ and $b$ denote the two stars, $j=1,\dots, N$ and $h$ represents respectively the nodes and hubs of the  stars. Hubs of the two stars are coupled diffusively to each other with inter-star coupling strength denoted by $\ep$. Thus the two coupled networks interacting through their highest degree nodes (hubs) to form a single network with two control parameters namely, the intra-star coupling strength $K$ and the inter-star coupling strength $\ep$. Therefore  it is natural to explore the global dynamics of the coupled stars as well as the dynamics of individual stars for different values of $K$ and $\ep$. To understand this transition, we plot the order parameter $R$ in $\ep-K$ parameter space. As shown in Fig.~\ref{fig6}, we observe that as the value of $\ep$ increases, the hysteresis width initially increases goes to a maximum value and then decreases if $\ep$ is increased further. This happens because the forward transition point changes with $\ep$, however, we do not observe any change in the backward transition point.

\begin{figure}
\centering
\scalebox{0.30}{\includegraphics[angle=270]{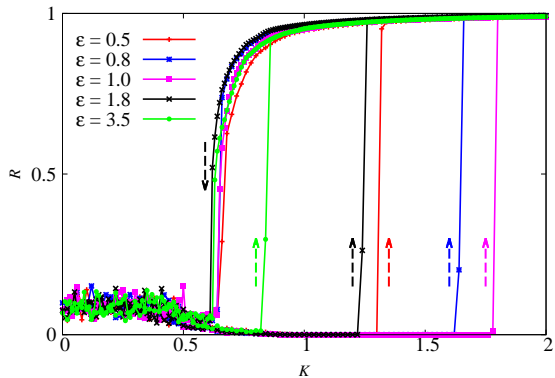}}
\caption{Variation of the global order parameter for Eq.~\ref{eq-dif} as a function of $K$ for different values of $\ep$.}
\label{fig7}
\end{figure}

\begin{figure}
\centering
\scalebox{0.30}{\includegraphics[angle=270]{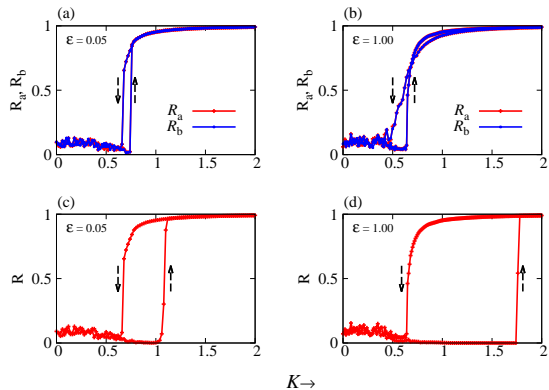}}
\caption{Upper panel shows variation of the order parameters ($R_a$ and $R_b$) for individual stars consisting with $N=100$ oscillators as a function of $K$ at (a) $\ep=0.05$, (b) $\ep=1$, whereas the lower panel describes the global order parameter ($R$) as a function of $K$ at (c) $\ep=0.05$, (d) $\ep=1$ for Eq.~\ref{eq-dif}. }
\label{fig8}
\end{figure}

 \begin{figure}
\centering
\scalebox{0.30}{\includegraphics[angle=270]{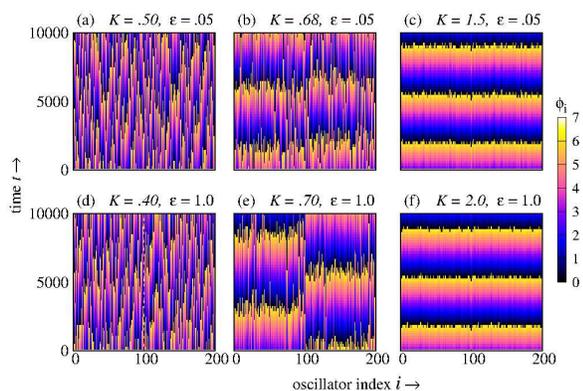}}
\caption{Upper panel represents the phase-time plots (Eq.~\ref{eq-dif}) for $\ep=0.05$ at (a) $K=0.5$, (b) $K=0.68$ and (c) $K=1.5$. In the lower panel shows the phase-time plots for $\ep=1$ at (d) $K=0.4$, (e) $K=0.7$ and (f) $K=2$. Individual star consists of $N=100$ oscillators.}
\label{fig9}
\end{figure}

To understand the dynamics described by the parameter space, we plot the global order parameter with respect to intra-star coupling strength ($K$) for different values of $\ep$. For individual stars (Eq.~\ref{eq-dif}) we consider $N=100$ oscillators. We observe that the networks show first order transition to synchrony accompanied by a hysteresis. In Fig.~\ref{fig7}, we plot the phase transition diagram for $\ep=0.5,0.8,1.0,1.8, 3.5$. Initially for $\ep<1$,  the hysteresis width increases with increasing $\ep$. As $\ep$ is increased beyond $\ep=1$, (say for $\ep=1.8, 3.5$) we observe that the hysteresis width decreases.  Interestingly, we observe that the backward transition point does not change with inter-star coupling. As the coupling $\ep$ is increased further, we observe that the hysteresis width saturates. From these results, we infer that inter-star coupling strengthens ES in case of interacting star networks and can be used as an additional means to control ES. Global transitions can be understood in terms of the dynamics at the level of individual stars.  In Fig.~\ref{fig8}, we plot the order parameters for layers $a$ and $b$ and compare it with the global order parameter for $\ep=0.05$ and $1.0$. As shown in Fig.~\ref{fig8} (a) and (c), keeping $\ep=0.05$ fixed, individual order parameters $R_a$ and $R_b$ show a first oder transition at much smaller values of $K$ as compared to the global order parameter $R$. Thus, we infer that although the two stars are in coherent state individually, the global dynamics is still desynchronized  ($R\sim0$). The two stars can be made to synchronize by increasing the intra-star coupling strength $K$ thereby increasing the hysteresis width for a fixed $\ep$. In Fig.\ref{fig9}, we plot phase-time diagrams for $\ep=0.05$ upper panel and $\ep=1$ in the lower panel. At $\ep=0.05$, the dynamics is incoherent  for $K=0.5$, which transforms to synchrony at the level of individual star at $K=0.68$ (\ie $R_a=R_b=1, R=0$) and finally at $K=1.5$, the two stars are synchronized. Similarly, the dynamics for $\ep=1$ is described in Fig.~\ref{fig8} (right panel) and the lower panel of  Fig.~\ref{fig9} shows the corresponding phase-time plots for Eq.~\ref{eq-dif} for different values of $K$. At $K=0.4$ the dynamics is desynchronized and all the three order parameters are zero ($R_a=R_b=R=0$) as shown in Fig.~\ref{fig8}(b) and (d) and the collective state is shown in Fig.~\ref{fig9}(d).  As the value of $K$ increases, the dynamics on the two stars become coherent $(R_a=R_b=1)$  but the global dynamics is still incoherent (c.f. Fig.~\ref{fig8}(b) and (d)) for which the collective dynamics is shown in (Fig.~\ref{fig9}(e)). Global synchrony is achieved as shown by the phase-time plots in Fig.~\ref{fig9}(f) for $K=2$ where $R_a=R_b=R=1$.

\subsection{Conjugate coupling}

In this section, we explore the idea of coupling between dissimilar variables, namely conjugate coupling \cite{karnatak-pre-2007, kim-prl-2005} between the hubs of the two stars. Originally, conjugate coupling was studied in context of amplitude death in case of two coupled Stuart-Landau oscillators \cite{karnatak-pre-2007}.  Dynamical effects of breaking rotational symmetry was studied in coupled SL oscillators and it was observed that depending upon the details of the system, the regime of oscillation death may occur or get suppressed. 
 Having studied the effect of coupling in similar variables, a natural question is to consider coupling in different variables.   Note that coupling in dissimilar variables destroys the symmetry in SL oscillators \cite{punetha-pre-2018}. Thus, two star networks with their hubs interacting via conjugate coupling, can be represented by equations of motion is given by

\begin{widetext}
\[
\begin{aligned}
\label{eq-conj}
\dot{z}_{a,j} &= \left[ (q+i\omega_{a,j}-|z_{a,j}|^2)z_{a,j}(t)+ K |\omega_{a,j}|  \left(z_{a,h}(t)-z_{a,j}(t)\right )\right], \nonumber\\
\dot{z}_{a,h} &= \alpha \bigl\{ (q+i\omega_{a,h}-|z_{a,h}|^2)z_{a,h}(t) +\frac{K |\omega_{a,h}|}{N} \sum_{j=1}^{N}  \left(z_{a,j}(t)-z_{a,h}(t)\right ) + \ep \left(\bar{z}_{b,h}(t)-z_{a,h}(t)\right)\bigr\}, \nonumber   \\
\dot{z}_{b,j} &= \left[ (q+i\omega_{b,j}-|z_{b,j}|^2)z_{b,j}(t) + K |\omega_{b,j}|  \left(z_{b,h}(t)-z_{b,j}(t)\right )\right], \nonumber \\
\dot{z}_{b,h} &= \alpha \bigl\{ (q+i\omega_{b,h}-|z_{b,h}|^2)z_{b,h}(t)  +\frac{K |\omega_{b,h}|}{N} \sum_{j=1}^{N}  \left(z_{b,j}(t)-z_{b,h}(t)\right ) + \ep \left(\bar{z}_{a,h}(t)-z_{b,h}(t)\right)\bigr\}, 
\end{aligned}
\]    
\end{widetext}

 \begin{figure}
\centering
\scalebox{0.35}{\includegraphics[angle=270]{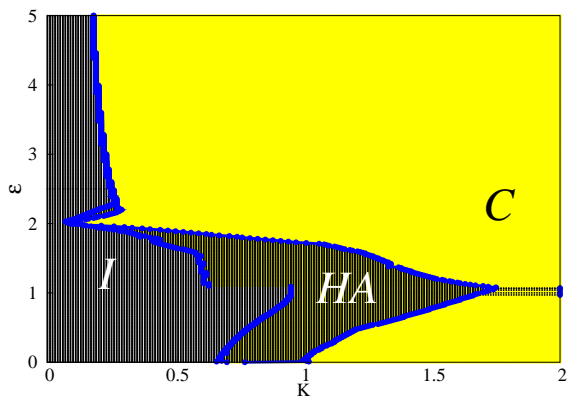}}
\caption{Variation of the global order parameter for different values of $\ep$. Black and yellow (light gray) regions identified by $I$ and $C$ indicates incoherent and coherent states. Region in between the blue lines represent the hysteresis area and is denoted as HA.}
\label{fig11}
\end{figure}

\begin{figure}
\centering
\scalebox{0.30}{\includegraphics[angle=270]{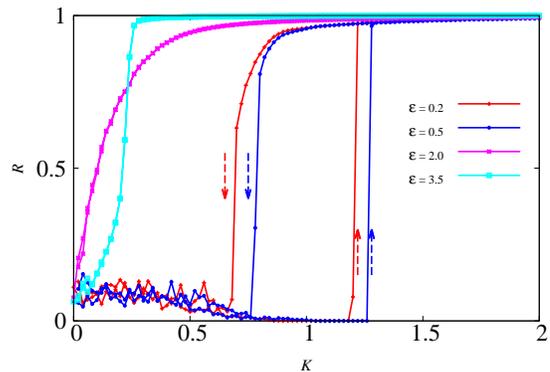}}
\caption{Variation of the global order parameter for Eq.~\ref{eq-conj} with $N=100$ oscillators for different values of $\ep$.}
\label{fig10}
\end{figure}

 To understand the features of the dynamics in general, we explore the dynamics through the global order parameter $R$ in $\ep-K$ parameter space as shown in Fig.~\ref{fig11}. The black, yellow (light gray) region and the region between blue curve represent  incoherent (I), coherent (C) and the hysteresis area  (HA), respectively. We observe that for small values of $\ep$, the transition to synchrony is a first order transition where the hysteresis area increases for smaller values of inter-star coupling. For large values of inter-star couplings ($\ep$), we observe that the transition becomes a second order transition. To describe it further, we plot the variation of the order parameter $R$ for four different values of $\ep$. For $\ep=0.2$ and $\ep=0.5$,  the transition is a first order transition with hysteresis as shown in Fig.~\ref{fig10}. However, as the value of $\ep$ is increased further, ES disappears and the transition to synchrony is a second order transition. This can be seen in Fig.~\ref{fig10} for the coupling values $\ep=2.0$ and $\ep=3.5$. Thus, the  conjugate coupling affects the global dynamics in two different ways depending upon the value of inter-star coupling $\ep$.  For relatively small couplings, the transition is first order in nature accompanied by a hysteresis, whereas for large values of $\ep$, the transition is second order in nature. To understand different transitions in the order parameter we explore the two cases separately.
 
 \begin{figure}
\centering
\scalebox{0.30}{\includegraphics[angle=270]{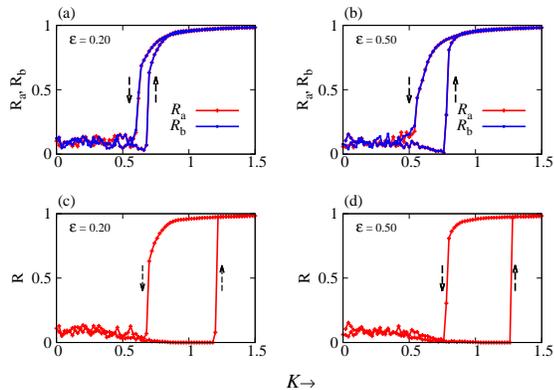}}
\caption{Plots to explain first order transition in the network of $N=100$ oscillators given by Eq.~\ref{eq-conj}. Variation of the order parameters $R_a$ and $R_b$ for individual stars for coupling values of (a) $\ep=0.2$ and (b) $\ep=0.5$. Identical plots for global order parameter for the coupling values (c) $\ep=0.2$ and (d) $\ep=0.5$.}
\label{fig12}
\end{figure}

\begin{figure}
\centering
\scalebox{0.30}{\includegraphics[angle=270]{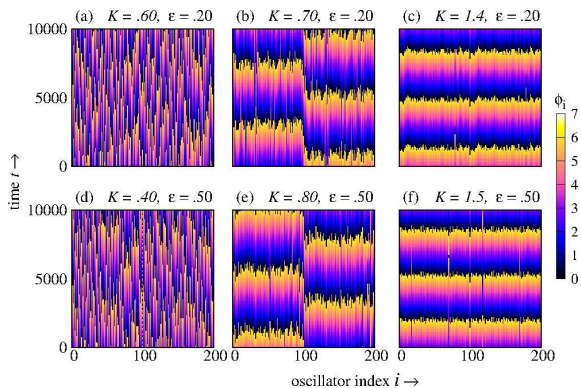}}
\caption{Upper panel represents the phase-time plots (Eq.~\ref{eq-conj}) for $\ep=0.2$ at (a) $K=0.6$, (b) $K=0.7$ and (c) $K=1.4$. Lower panel shows the phase-time plots for $\ep=0.5$ at (d)$K=0.4$, (e) $K=0.8$ and (f) $K=1.5$. Each star network consists with $N=100$ oscillators.}
\label{fig13}
\end{figure}

For $\ep=0.2$ and $\ep=0.5$, where the order parameter changes abruptly, we plot the order parameters ($R_a$,  $R_b$ and $R$) for individual stars  as shown in Figs.~\ref{fig12} (a) and (c). At $\ep=0.2$,  we observe that the order parameters are zero for $K=0.6$ for which the phase-time plots are shown in Fig.~\ref{fig13}(a). Note that the for $K=0.7$, the order parameters $R_a,R_b \rightarrow 1$ and $R=0$ as shown in Figs.~\ref{fig12} (a) and (c) indicating that the individual stars are synchronized where as the global dynamics is desynchronized. Collective states for this transition is shown in Fig.~\ref{fig13}(b).  Global dynamics becomes coherent at $K=1.4$, where ($R_a=R_b=R=1$) and the collective dynamics is plotted in Fig.~\ref{fig13}(c). Similarly, for $\ep=0.5$, our observations are qualitatively same. Order parameters are plotted in Figs.~\ref{fig12} (b) and (d) and the collective states for $K=0.4, K=0.8$ and $K=1.5$ are plotted in Figs.~\ref{fig13}(d), (e) and (f) respectively. Thus, we conclude that in case of conjugate coupling, the transition in individual stars is a first order transition, but the dynamics on these stars may be incoherent w.r.t each other and hence the global order parameter $R$ remains zero upto a very large value of intra-star coupling (K). This case is identical to the one obtained in case of direct coupling.

\begin{figure}
\centering
\scalebox{0.30}{\includegraphics[angle=270]{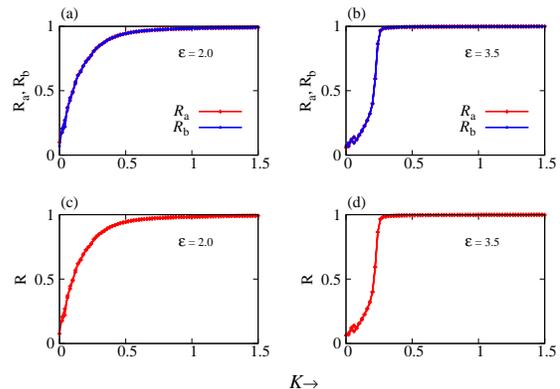}}
\caption{Plots to explain second order transition in the network given by Eq.~\ref{eq-conj}. Variation of the order parameters $R_a$ and $R_b$ for individual stars for coupling values of (a) $\ep=0.2$ and (b) $\ep=0.5$. Identical plots for global order parameter for the coupling values (c) $\ep=0.2$ and (d) $\ep=0.5$.}
\label{fig14}
\end{figure}

On the contrary, in case of conjugate coupling,  we also observe that the global order parameter shows a second order transition for large values of $\ep~(\ep>2.00)$  as shown in Figs.~\ref{fig11}. To understand the transitions in the individual stars, we plot the order parameters $R_a$ and $R_b$ at $\ep=2$ and $\ep=3.5$ as shown in Fig.~\ref{fig14}. At $\ep=2$, the transition to synchrony is continuous as shown by the order parameters $R_a$ and $R_b$ in Fig.~\ref{fig14}(a). The global order parameter, $R$ also changes in a continuous manner as shown in Fig.~\ref{fig14}(c). Similarly, at $\ep=3.5$, we observe that the transition in the two stars is second order in nature as shown in Fig.\ref{fig14}(b). Similarly, the global order parameter $R$ changes continuously at $\ep=3.5$ as shown in Fig.~\ref{fig14}(d).  Thus, in case of second order transitions, the global dynamics as well as dynamics on the individual star becomes coherent at the same value of intra-star coupling $K$.

\section{Summary} 
\label{summary}
In conclusion, we have presented a general framework to obtain ES in an ensemble of Stuart-Landau oscillators.  By introducing frequency weighted coupling and time scale variation, it was possible to induce degree-frequency correlations in a star network. Addition of time scale variation allows us to obtain ES with a well defined hysteresis. The study was further extended to a case of two star networks coupled through their hubs. We explore the effect of coupling strength on the global dynamics in presence of direct diffusive coupling and conjugate coupling. Global dynamics in presence of these couplings is explained in terms of the dynamics on individual stars. In case of direct coupling, the hysteresis width increases with increasing inter-star coupling and then saturates for large values. Results for conjugate coupling are same as that of direct coupling for small values of coupling strengths. For large values of couplings, the hysteresis width decreases and the transition becomes a second order. This is shown by plotting the global order parameter which was found to increase continuously with increasing inter-star coupling. Thus, one can couple two (or more) star networks to design a network with desired global properties. The techniques outlined here are very general and robust and can be extended to explore ES by considering  $(i)$ more complicated dynamics on the nodes, $(ii)$ interactions of three or  more stars and $(iii)$ interaction of nonidentical star networks.

\end{document}